# Electron Collimation in Twisted Bilayer Graphene via Gate-defined Moiré Barriers

Wei Ren[1], Xi Zhang[1], Ziyan Zhu[2], Moosa Khan[1], Kenji Watanabe[3], Takashi Taniguchi[4], Efthimios Kaxiras[5,6], Mitchell Luskin[7], Ke Wang[1*]

[1]*School of Physics and Astronomy, University of Minnesota, Minneapolis, Minnesota 55455, USA*

[2]*Stanford Institute for Materials and Energy Sciences, SLAC National Accelerator Laboratory, Menlo Park, CA 94025, USA*

[3]*Research Center for Electronic and Optical Materials, National Institute for Materials Science, 1-1 Namiki, Tsukuba 305-0044, Japan*

[4]*Research Center for Materials Nanoarchitectonics, National Institute for Materials Science, 1-1 Namiki, Tsukuba 305-0044, Japan*

[5]*Department of Physics, Harvard University, Cambridge, Massachusetts 02138, USA*

[6]*John A. Paulson School of Engineering and Applied Sciences, Harvard University, Cambridge, Massachusetts 02138*

[7]*School of Mathematics, University of Minnesota, Minneapolis, Minnesota 55455, USA*

**Electron collimation via a graphene pn-junction allows electrostatic control of ballistic electron trajectories akin to that of an optical circuit. Similar manipulation of novel correlated electronic phases in twisted-bilayer graphene (tBLG) can provide additional probes to the underlying physics and device components towards advanced quantum electronics. In this work, we demonstrate collimation of the electron flow via gate-defined moiré barriers in a tBLG device, utilizing the band-insulator gap of the moiré superlattice. A single junction can be tuned to host a chosen combination of conventional pseudo barrier and moiré tunnel barriers, from which we demonstrate improved collimation efficiency. By measuring transport through two consecutive moiré collimators separated by 1 um, we demonstrate evidence of electron collimation in tBLG in the presence of realistic twist-angle inhomogeneity.**

A massless relativistic particle can tunnel through a barrier with 100% transmission, independent of the barrier height and width, a phenomenon known as Klein paradox. Analogous behavior has been observed in graphene thanks to the linear energy-momentum dispersion relationship for electrons, akin to a massless relativistic particle. In the case of electrons in graphene, there exists a pseudo-barrier defined at the interface of a pn-junction: ballistic carriers along paths perpendicular to the pn-junction undergo perfect transmission, whereas carriers astray from it get reflected with a probability that increases exponentially with incident angle [1–8]. The strong angle dependence of Klein tunneling allows for collimation of ballistic electron trajectories, or the generation of electronic plane waves. This can serve as the basis for electron-optics, in which electrons are manipulated to form electronic analogies of optical circuits, whose coherent interference can be utilized for novel quantum electronics.

The presence of flat bands in twisted bi- and multi-layer graphene leads to electron localization and correlations [9–11]; this allows the emergence of novel quantum transport phenomena, such as superconductivity [12–23], correlated insulator states [21–35] and magnetism [21,24,35–37]. The collimation of correlated ballistic current can be used for more versatile manipulation of the novel quantum coherent phases towards advanced quantum electronics, including applications in spintronics, valleytronics, and guided Josephson currents; in addition, it may provide flexible experimental knobs for investigating the microscopic mechanism of the rich underlying physics of correlated electron states.

There remain two major experimental challenges in realizing these possibilities: first and foremost, the collimation efficiency becomes better with a narrower and higher tunnel barrier [1]. Realistically, the electrostatically defined pn-junction is at least as wide as the gate separation (~50 nm), and the effective pseudo-barrier height is limited by the absence of a true bandgap. As a result, a portion of the electrons with small but finite incident angle undergo Veselago refraction instead of being reflected, resulting in transmission of these uncollimated electrons that in turn compromises the overall collimation efficiency [38–42]. The second challenge has to do with the fact that the ballistic electron trajectory in twisted-bilayers may be curved over long distance, in contrast to that in monolayer graphene, due to twist-angle inhomogeneity from local atomic reconstruction which is present in real samples. After passing through the collimator [43–46], electron current can gradually go astray from its intended path, making it harder to stay collimated over long distance.

In this work, we demonstrate collimation of the electron flow in tBLG via gate-defined moiré barriers. Utilizing the moiré band-insulator gap, we show that a narrow (on the order of 10 nm) and a true tunnel barrier can be electrostatically defined to improve the collimation efficiency on top of the state-of-the-art pn-pseudo barriers (See Supporting Information section S1, S5). By measuring the transport through two consecutive moiré collimators separated by 1 um, we demonstrate evidence of electron collimation in tBLG in the context of the additional challenges presented by twist-angle inhomogeneity.

Figure 1a shows the schematic image of a single moiré barrier. Two pieces of monolayer graphene (MLG) are consecutively picked up with a relative twist angle of ~1.05°, where a band insulator gap (Figure 1d) of ~ 23 meV is expected. The stack is encapsulated by hBN and transferred onto two pre-patterned local bottom gates (hereby referred to as $G_1$ and $G_2$), whose topography is confirmed by atomic force microscopy to be atomically flat (Figure 1e, also see SI section S1). This is a crucial experimental design to avoid additional atomic strain, as local twist-angle homogeneity in the region between two gates is essential for a spatially uniform moiré tunnel barrier. The two bottom gates tune the carrier density (and the filling factor) in regions directly above them, and the carrier density changes linearly in tBLG across the region in between the ~100 nm gate separation.

Figure 1f shows measured 4-probe resistance as a function of the filling factors, $\nu_1$ and $\nu_2$, of the region on top of $G_1$ and $G_2$, respectively. Local high resistance is observed along three horizontal and vertical lines, when either one of the two regions reach precisely $\nu = -4, 0$ or $4$ as expected from the relatively high resistance of the charge neutrality point and band-insulator states. Here the filling factor is defined as $\nu = n/(n_S/4)$, where $n$ is the carrier density in tBLG, and $n_S$ is the carrier density corresponding to 4 electrons per moiré unit cell. The experiments are conducted at an electron temperature $\geq 4$ K, at which correlated insulating states at $\nu = \pm 2$ have not been observed. This ensures that the measured electron collimation is due to tunnel barriers defined by band-insulator states ($\nu = \pm 4$) and the PN junction ($\nu = 0$). As a result, it can be interpreted solely within the frame of single-particle Klein physics.

When neither of the two filling factors $\nu_1$ and $\nu_2$ equals exactly $-4, 0$ or $4$, a series of resistive or insulating regions can exist in between the gate separation, depending on whether $\nu = -4, 0$ or $4$ is found in between the value of $\nu_1$ and $\nu_2$. For example (Figure 1b), at $\nu_1 = 5$ and $\nu_2 = 3$, as the filling factor in the region above the gate separation changes from $\nu = 5$ to $\nu = 3$, a narrow section of it reaches $\nu = 4$ band insulator states and serve as a narrow tunnel barrier, which we refer to as a $\nu = 4$ moiré barrier. The width of the barrier can be tuned to be narrower (wider) with an increase (decrease) in $|\nu_1 - \nu_2|$, and multiple tunnel barriers can co-exist to further improve collimation efficiency (See Supporting Information section S1). For example (Figure 1c), for $\nu_1 = 5$ and $\nu_2 = -5$, three narrow barriers corresponding to $\nu = -4$, $\nu = 0$ and $\nu = 4$, respectively, exist in series at the junction. This allows three consecutive collimation processes from two moiré barriers ($\nu = \pm 4$) and a pn-barrier ($\nu = 0$).

Strictly speaking, the tunneling across the pn-and moiré barriers in tBLG is not a conventional Klein tunneling process. The former relies on a pseudo-barrier while the latter lacks the electron-hole symmetry [1]. Yet for as long as the tunnel probability shows strong angle-dependence, collimation can be achieved. The barriers selectively reflect electrons with finite incident angle, and those successfully transmitted through follows a current path perpendicular to the barrier. At the ballistic limit, the resistance of the junction $R_J$ measures the number of electrons it reflects (or backscatters), and can be extracted by symmetrizing the measured 4-probe resistance (to eliminate gate-dependent contact resistance and bulk graphene resistance) following the previously-established methods [8] (See Supporting Information section S2 for the detailed method).

Figure 2a shows the extracted junction resistance $R_J$ as a function of fillings factors $\nu_1$ and $\nu_2$. 16 sections separated by $\nu_1, \nu_2 = -4, 0, 4$ lines are clearly visible, each corresponding to an arbitrary combination of $\nu = -4$, $\nu = 0$ and $\nu = 4$ barriers found (or absent) in the gate-defined junction. For ease of referencing, we label the column numbers A, B, C and D for $\nu_1 < -4$, $-4 < \nu_1 < 0$, $0 < \nu_1 < 4$ and $\nu_1 > 4$, respectively, and row number a, b, c and d for $\nu_2 < -4$, $-4 < \nu_2 < 0$, $0 < \nu_2 < 4$ and $\nu_2 > 4$, respectively. The number and type of tunnel barriers existing in the junction depends on whether $\nu = -4, 0$ and/or 4 are found in between $\nu_1$ and $\nu_2$. For example, the domain Ca corresponds to $0 < \nu_1 < 4$ and $\nu_2 < -4$, in between which $\nu = -4, 0$ are found. Therefore, a junction consisting of a pn-barrier and a $\nu = -4$ moiré barrier is expected. Similarly, a junction in Aa, Bb, Cc and Dd contains no tunnel barriers, while that in Ad and Da contains all three in series.

A previously established method for characterizing the collimation efficiency [8] compares the junction resistance of two co-existing barriers in series ($B_{12}$), to that with just one at a time ($B_1$ or $B_2$). In the limit of perfect collimation by barrier $B_1$, charge carriers will experience no reflection at $B_2$ whether $B_2$ is established or not, leading to $R_{B_{12}} = R_{B_1}$. In contrast, if collimation is completely absent, we expect $R_{B_{12}} = R_{B_1} + R_{B_2}$, a trivial consequence of two resistors in series. Previous work with state-of-the-art pn-junctions have demonstrated $R_{B_1} + R_{B_2} > R_{B_{12}} > \max(R_{B_1}, R_{B_2})$, with ~ 40% and ~25% difference between neighboring terms [8], and an estimated collimation efficiency of ~30%.

We first examine the pair of $\nu = 0$ (pn) and $\nu = 4$ (moiré) barriers along a 1D cut (Figure 2b) taken across Bc→Cc→Cd→Bd (red) in Figure 2a, where the junction contains tunnel barriers of Bc:[$\nu = 0$] → Cc:[none] → Cd:[$\nu = 4$] → Bd:[$\nu = 0, 4$] (both, in series). The observation of $R_{\nu=0} + R_{\nu=4} > R_{\nu=0,4} \approx R_{\nu=4}$ implies nearly perfect (some) collimation from $\nu = 4$ moiré barrier ($\nu = 0$ pn-barrier). The similar junction resistance observed in Cd and Bd, implies that electrons with finite incident angle get efficiently reflected by $\nu = 0$ barrier alone (Cd). The resulting electrons are highly collimated and can fully transmit through the downstream pn-barrier (added in Bd), without additional reflection to increase the junction resistance further. A 1D cut along column D (yellow) in Figure 2a adds one barrier at a time, with junction consisting of barriers from Dd:[none] → Dc: [$\nu = 4$] → Db: → [$\nu = 0, 4$] → Da: [$\nu = -4, 0, 4$]. The junction resistance reaches its maximum value as soon as the $\nu = 4$ barrier is added and single-handedly reflects all uncollimated electrons. Subsequently added $\nu = 0$ and $\nu = -4$ barriers become nearly invisible to the collimated electrons downstream of the $\nu = 4$ moiré barriers and are thus incapable of raising the junction resistance any further.

In contrast, the collimation efficiency of the $\nu = -4$ moiré barrier is observed to be much weaker (See Supporting Information section S5 for more detailed discussion). Figure 2d shows a 1D cut across Cb → Bb → Ba → Ca (blue). While $R_{\nu=0} + R_{\nu=-4} > R_{\nu=-4,0}$ (see SI section S4 for more details), the collimation efficiency of the $\nu = -4$ moiré barrier seems to be no better than that of the conventional pn-barrier. A cut across column A (Figure 2e) along Aa→Ab→Ac→Ad (green) confirms similarly. After the $\nu = -4$ barrier is added (Ab), some of the uncollimated electrons are reflected by an additional pn-barrier (Ac). Even the combined collimation efficiency of the two is not perfect since the final addition of the strong $\nu = 4$ moiré collimator still helps reflect electrons that managed to escape through the $\nu = -4, 0$ barriers.

Multiple devices qualitatively reproduced the results (See Supporting Information section S4), while the quantitative collimation efficiency for the moiré barrier is extremely sensitive to homogeneity of local twist angles and the electrostatic profile at the junction, with realistic device-to-device variation. For comparison, we measured a second junction ($J_2$), electrostatically defined on the same tBLG sample, 1 µm away from the first junction ($J_1$). Figure 3a shows the junction resistance as a function of $\nu_1$ and $\nu_2$. The $\nu_1, \nu_2 = -4, \ 0, \ 4$ local resistance peaks separating the 16 sections are found at nearly the exact same (See Supporting Information section S4) carrier densities as that of $J_1$, showing that the overall twist-angle has not migrated over the 1 µm separation between the two devices. 1D cuts (Figure 3b, 3c) through Column A and row d, implies neither of the two ($\nu = -4$ and $\nu = 4$) moiré barriers are enough to reach a high level of collimation by itself. However, the junction resistance with two consecutive barriers (Ac: $\nu = 0, 4$ or Bd: $\nu = -4, 0$), is much more similar to that of the "full-house" (Ad: $\nu = -4, 0, 4$). This shows that effective collimation can be achieved with two tunnel barriers in the junction, and the addition of the third and final barrier no longer helps and thus is unnecessary.

The above measurements demonstrate that the gate-defined moiré barrier significantly improves the collimation efficiency, either single-handedly or in collaboration with a conventional pn-barrier. However, such measurement is done by tracking the ballistic electron trajectory over a distance no longer than the 100 nm gate separation. The ballistic electron flow across longer distances may not follow a straight path (like that in monolayer graphene), due to inhomogeneity of the local twist-angle in the sample, and therefore the spatially varying electronic structure.

To characterize long-distance collimation efficiency in tBLG, we measure overall resistance of the entire device consisting of the two junctions ($J_1$ and $J_2$ mentioned above) separated by 1 µm distance, with three local gates that can independently tune the carrier density directly above. Each junction can contain one of the barriers ($\nu = -4, 0$, or $4$) or a combination of them. Three sets of measurements are performed with the same circuits but different gate configurations (Fig. 4b) and are symmetrized to eliminate the contribution from bulk graphene resistance and contact resistance. The resistance of right (left) junction $J_1$ ($J_2$) as a function of adjacent bulk carrier densities, is measured by configuring the center gate voltage to be the same as the left (right) gate, to ensure that the other junction is absent. The junction resistance $J_{12}$ is measured when both junctions are simultaneously defined, with left and right gate voltage tuned to the same value, and therefore the same barrier combinations are expected in each junction throughout the measurement. While it is possible to configure the two junctions differently, the extraction of $J_{12}$ with three independent gate voltages requires symmetrizing 3-dimensional data sets, which is unrealistic.

The difference of $\Delta R = R_{J_1} + R_{J_2} - R_{J_{12}}$ is plotted to characterize the signature of collimation (Figure 4a), with $\Delta R = 0$ corresponding to the trivial scenario of two resistors in series. $\Delta R = 0$ is measured when collimation is completely missing in sections Aa, Bb, Cc and Dd, as expected. A positive $\Delta R$ is observed in section Cd and Dc, when both junctions consist of only $\nu = 4$ moiré barriers, demonstrating a signature of electron collimation over 1 µm distance in tBLG. Adding more barriers to each junction improves (or at least does not comprise) the collimation efficiency of each junction based on previous characterization, yet the resulting $\Delta R$ decreases to nearly zero. This implies that a highly collimated electron beam from one junction is unsuccessful in transmitting through the other.

We attribute this observation to curved electron ballistic paths due to local twist-angle inhomogeneity. To elaborate on this, we plot the electron flow at the limit of nearly perfect collimation by each junction (Figure 4c), where the angle-dependence is close to a d-function that reflects any electron other than the one incident strictly perpendicular to the barrier. The curvature of the collimated electron path over the 1 µm distance results in reflections of electrons with finite incident angle, instead of 100% transmission. Significant backscattering occurs at both junctions (akin to resistors in series), despite the highly guided electron path, thus eliminating the expected logic between the two collimators. As a result, $\Delta R > 0$ is measured when there is effective but not perfect collimation (Figure 4d). Electrons passing through $J_2$ have a narrow but still finite angle distribution, permitting electron paths through $J_2$ with small incident angle. A selection of those electrons will arrive perpendicularly to $J_1$ after traveling a curved ballistic path, successfully transmitting through $J_1$ with ease, and thus contributing to $\Delta R$. Nonetheless, the results experimentally confirm the expected curved ballistic electron path in tBLG over long distance, and electrostatic manipulation over it. Future experiments with scanning gate microscopy can further confirm the microscopic details of the electron trajectories. In applications where a straighter path is desired, a series of gate-defined junctions with smaller separations can be implemented.

In conclusion, we demonstrate electrostatically defined junctions in tBLG, consisting of arbitrary combinations of gate-defined tunnel barriers from two band-insulator gaps and conventional pn-barriers. We demonstrate that the collimation efficiency of electrons can be improved with a moiré barrier and observe the signature of long-distance collimation in the presence of realistic twist-angle inhomogeneity. Our work provides new insights in electron transport in moiré superlattices, and new approaches in manipulating the exotic quantum transport phenomena, both as additional probes to rich underlying physics and as components towards advanced quantum electronic circuits.

We thank Alex Kamenev and Dan Dahlberg for useful discussions. This work was supported by NSF DMREF Award DMR-1922165 and Award DMR-1922172, ARO MURI Grant No. W911NF-14-1-0247 and Simons Foundation Award no. 896626. The development of atomically-flat and strain-free gate-defined nanostructure was supported by NSF CAREER Award 1944498. Nanofabrication was conducted in the Minnesota Nano Center, which is supported by the National Science Foundation through the National Nano Coordinated Infrastructure Network, Award Number NNCI -1542202. Z.Z. is supported by a Stanford Science fellowship. Portions of the hexagonal boron nitride material used in this work were provided by K.W and T. T. K.W. and T.T. acknowledge support from the JSPS KAKENHI (Grant Numbers 20H00354, 21H05233 and 23H02052) and World Premier International Research Center Initiative (WPI), MEXT, Japan.

**Sample Preparation and Device Fabrication.** A pair of parallel metal gates ($G_1$ and $G_2$, with a 120 nm separation between them) consisting of Cr/Pd-Au alloy (1 nm / 7 nm) is deposited on a $SiO_2$ (285 nm) / Si (doped) substrate. The Pd-Au alloy (40% Pd / 60% Au) is chosen to reduce the surface roughness when it is compared to conventional pure Au deposition. The gates are subsequently annealed in a high-vacuum environment at 350 ℃ for 10 minutes to remove surface residue.

The tBLG devices are made by the 'cut and tear' method [22]. First, hBN and monolayer graphene flakes are exfoliated [47] and characterized by the atomic force microscope (AFM) to be atomically clean. A single piece of monolayer graphene was cut into two individual pieces along the same lattice orientation by the atomic force microscope (AFM). With the help of a poly (bisphenol A carbonate) (PC) and polydimethylsiloxane (PDMS) stamp on a glass slide [48], we pick up the first piece of hBN. Then the two pieces of the precut graphene are picked up consecutively, with a relative twist angle $\theta = 1.05°$ between them. The second piece of hBN is picked up to encapsulate the tBLG and the whole stack is transferred onto the pre-deposited bottom metal [49]. After the PC residue on the top hBN surface is cleaned by chloroform, acetone and isopropanol, the Cr/Pd/Au (1 nm/5 nm/180 nm) metal contacts (serve as source, drain and voltage probes) are added to the sample by e-beam lithography, plasma etching and e-beam evaporation processes [50]. Finally, a Hall-bar shaped bubble-free region is defined by e-beam lithography and plasma etching.

**Electrical Transport Measurements.** Experiments are performed in a Montana fridge at a base temperature of ~4 K. All data are collected by a standard lock-in amplifier with an alternating current excitation of 10 nA at 17.777778 Hz applied through the device unless otherwise specified. Yokogawa and Keithley DC voltage sources are used to apply voltages on the local metal gates and the global Si backgated, respectively.

[\*]kewang@umn.edu

[1] M. I. Katsnelson, K. S. Novoselov, and A. K. Geim, *Chiral Tunnelling and the Klein Paradox in Graphene*, Nat Phys **2**, 620 (2006).

[2] N. Stander, B. Huard, and D. Goldhaber-Gordon, *Evidence for Klein Tunneling in Graphene Pn-Junctions*, Phys Rev Lett **102**, 026807 (2009).

[3] C. W. J. Beenakker, *Colloquium: Andreev Reflection and Klein Tunneling in Graphene*, Rev. Mod. Phys. **80**, 1337 (2008).

[4] A. F. Young and P. Kim, *Quantum Interference and Klein Tunnelling in Graphene Heterojunctions*, Nat Phys **5**, 222 (2009).

[5] Q. Wilmart, S. Berrada, D. Torrin, V. H. Nguyen, G. Fève, J.-M. Berroir, P. Dollfus, and B. Plaçais, *A Klein-Tunneling Transistor with Ballistic Graphene*, 2D Mater. **1**, 011006 (2014).

[6] S. Chen, Z. Han, M. M. Elahi, K. M. Habib, L. Wang, B. Wen, Y. Gao, T. Taniguchi, K. Watanabe, and J. Hone, *Electron Optics with Pn Junctions in Ballistic Graphene*, Science **353**, 1522 (2016).

[7] M.-H. Liu, C. Gorini, and K. Richter, *Creating and Steering Highly Directional Electron Beams in Graphene*, Phys Rev Lett **118**, 066801 (2017).

[8] K. Wang, M. M. Elahi, L. Wang, K. M. Habib, T. Taniguchi, K. Watanabe, J. Hone, A. W. Ghosh, G.-H. Lee, and P. Kim, *Graphene Transistor Based on Tunable Dirac Fermion Optics*, Proc. Natl. Acad. Sci. **116**, 6575 (2019).

[9] R. Bistritzer and A. H. MacDonald, *Moiré Bands in Twisted Double-Layer Graphene*, Proc. Natl. Acad. Sci. **108**, 12233 (2011).

[10] Z. Zhu, S. Carr, D. Massatt, M. Luskin, and E. Kaxiras, *Twisted Trilayer Graphene: A Precisely Tunable Platform for Correlated Electrons*, Phys. Rev. Lett. **125**, 116404 (2020).

[11] Z. Ma, S. Li, Y.-W. Zheng, M.-M. Xiao, H. Jiang, J.-H. Gao, and X. C. Xie, *Topological Flat Bands in Twisted Trilayer Graphene*, Sci. Bull. **66**, 18 (2021).

[12] Y. Cao, V. Fatemi, S. Fang, K. Watanabe, T. Taniguchi, E. Kaxiras, and P. Jarillo-Herrero, *Unconventional Superconductivity in Magic-Angle Graphene Superlattices*, Nature **556**, 7699 (2018).

[13] M. Yankowitz, S. Chen, H. Polshyn, Y. Zhang, K. Watanabe, T. Taniguchi, D. Graf, A. F. Young, and C. R. Dean, *Tuning Superconductivity in Twisted Bilayer Graphene*, Science **363**, 1059 (2019).

[14] J. M. Park, Y. Cao, K. Watanabe, T. Taniguchi, and P. Jarillo-Herrero, *Tunable Strongly Coupled Superconductivity in Magic-Angle Twisted Trilayer Graphene*, Nature **590**, 7845 (2021).

[15] Z. Hao, A. M. Zimmerman, P. Ledwith, E. Khalaf, D. H. Najafabadi, K. Watanabe, T. Taniguchi, A. Vishwanath, and P. Kim, *Electric Field–Tunable Superconductivity in Alternating-Twist Magic-Angle Trilayer Graphene*, Science **371**, 1133 (2021).

[16] X. Liu, N. J. Zhang, K. Watanabe, T. Taniguchi, and J. I. A. Li, *Isospin Order in Superconducting Magic-Angle Twisted Trilayer Graphene*, Nat Phys **18**, 5 (2022).

[17] H. Kim, Y. Choi, C. Lewandowski, A. Thomson, Y. Zhang, R. Polski, K. Watanabe, T. Taniguchi, J. Alicea, and S. Nadj-Perge, *Evidence for Unconventional Superconductivity in Twisted Trilayer Graphene*, Nature **606**, 7914 (2022).

[18] J. M. Park, Y. Cao, L.-Q. Xia, S. Sun, K. Watanabe, T. Taniguchi, and P. Jarillo-Herrero, *Robust Superconductivity in Magic-Angle Multilayer Graphene Family*, Nat. Mater. **21**, 8 (2022).

[19] Y. Zhang et al., *Promotion of Superconductivity in Magic-Angle Graphene Multilayers*, Science **377**, 1538 (2022).

[20] A. Uri et al., *Superconductivity and Strong Interactions in a Tunable Moiré Quasicrystal*, Nature **620**, 7975 (2023).

[21] X. Lu et al., *Superconductors, Orbital Magnets and Correlated States in Magic-Angle Bilayer Graphene*, Nature **574**, 7780 (2019).

[22] Y. Saito, J. Ge, K. Watanabe, T. Taniguchi, and A. F. Young, *Independent Superconductors and Correlated Insulators in Twisted Bilayer Graphene*, Nat. Phys. **16**, 9 (2020).

[23] X. Zhang, K.-T. Tsai, Z. Zhu, W. Ren, Y. Luo, S. Carr, M. Luskin, E. Kaxiras, and K. Wang, *Correlated Insulating States and Transport Signature of Superconductivity in Twisted Trilayer Graphene Superlattices*, Phys. Rev. Lett. **127**, 166802 (2021).

[24] Y. Saito, J. Ge, L. Rademaker, K. Watanabe, T. Taniguchi, D. A. Abanin, and A. F. Young, *Hofstadter Subband Ferromagnetism and Symmetry-Broken Chern Insulators in Twisted Bilayer Graphene*, Nat. Phys. **17**, 4 (2021).

[25] Y. Cao et al., *Correlated Insulator Behaviour at Half-Filling in Magic-Angle Graphene Superlattices*, Nature **556**, 7699 (2018).

[26] K. P. Nuckolls, M. Oh, D. Wong, B. Lian, K. Watanabe, T. Taniguchi, B. A. Bernevig, and A. Yazdani, *Strongly Correlated Chern Insulators in Magic-Angle Twisted Bilayer Graphene*, Nature **588**, 7839 (2020).

[27] I. Das, X. Lu, J. Herzog-Arbeitman, Z.-D. Song, K. Watanabe, T. Taniguchi, B. A. Bernevig, and D. K. Efetov, *Symmetry-Broken Chern Insulators and Rashba-like Landau-Level Crossings in Magic-Angle Bilayer Graphene*, Nat. Phys. **17**, 6 (2021).

[28] S. Wu, Z. Zhang, K. Watanabe, T. Taniguchi, and E. Y. Andrei, *Chern Insulators, van Hove Singularities and Topological Flat Bands in Magic-Angle Twisted Bilayer Graphene*, Nat. Mater. **20**, 4 (2021).

[29] S. Chen et al., *Electrically Tunable Correlated and Topological States in Twisted Monolayer–Bilayer Graphene*, Nat. Phys. **17**, 3 (2021).

[30] S. Xu et al., *Tunable van Hove Singularities and Correlated States in Twisted Monolayer–Bilayer Graphene*, Nat. Phys. **17**, 5 (2021).

[31] G. W. Burg, J. Zhu, T. Taniguchi, K. Watanabe, A. H. MacDonald, and E. Tutuc, *Correlated Insulating States in Twisted Double Bilayer Graphene*, Phys. Rev. Lett. **123**, 197702 (2019).

[32] Y. Cao, D. Rodan-Legrain, O. Rubies-Bigorda, J. M. Park, K. Watanabe, T. Taniguchi, and P. Jarillo-Herrero, *Tunable Correlated States and Spin-Polarized Phases in Twisted Bilayer–Bilayer Graphene*, Nature **583**, 7815 (2020).

[33] M. He, Y. Li, J. Cai, Y. Liu, K. Watanabe, T. Taniguchi, X. Xu, and M. Yankowitz, *Symmetry Breaking in Twisted Double Bilayer Graphene*, Nat. Phys. **17**, 1 (2021).

[34] C. Shen et al., *Correlated States in Twisted Double Bilayer Graphene*, Nat. Phys. **16**, 5 (2020).

[35] M. He, Y.-H. Zhang, Y. Li, Z. Fei, K. Watanabe, T. Taniguchi, X. Xu, and M. Yankowitz, *Competing Correlated States and Abundant Orbital Magnetism in Twisted Monolayer-Bilayer Graphene*, Nat Commun **12**, 1 (2021).

[36] A. L. Sharpe, E. J. Fox, A. W. Barnard, J. Finney, K. Watanabe, T. Taniguchi, M. A. Kastner, and D. Goldhaber-Gordon, *Emergent Ferromagnetism near Three-Quarters Filling in Twisted Bilayer Graphene*, Science **365**, 605 (2019).

[37] H. Polshyn et al., *Electrical Switching of Magnetic Order in an Orbital Chern Insulator*, Nature **588**, 7836 (2020).

[38] X. Zhang et al., *Gate-Tunable Veselago Interference in a Bipolar Graphene Microcavity*, Nat Commun **13**, 1 (2022).

[39] S. P. Milovanović, D. Moldovan, and F. M. Peeters, *Veselago Lensing in Graphene with a Pn Junction: Classical versus Quantum Effects*, J. Appl. Phys. **118**, 154308 (2015).

[40] S. P. Milovanović, M. Ramezani Masir, and F. M. Peeters, *Magnetic Electron Focusing and Tuning of the Electron Current with a Pn-Junction*, J. Appl. Phys. **115**, 043719 (2014).

[41] V. V. Cheianov, V. Fal'ko, and B. L. Altshuler, *The Focusing of Electron Flow and a Veselago Lens in Graphene Pn Junctions*, Science **315**, 1252 (2007).

[42] Y. Jiang, J. Mao, D. Moldovan, M. R. Masir, G. Li, K. Watanabe, T. Taniguchi, F. M. Peeters, and E. Y. Andrei, *Tuning a Circular Pn-Junction in Graphene from Quantum Confinement to Optical Guiding*, Nat. Nanotechnol. **12**, 11 (2017).

[43] B. Padhi, A. Tiwari, T. Neupert, and S. Ryu, *Transport across Twist Angle Domains in Moiré Graphene*, Phys. Rev. Res. **2**, 033458 (2020).

[44] S. Joy, S. Khalid, and B. Skinner, *Transparent Mirror Effect in Twist-Angle-Disordered Bilayer Graphene*, Phys. Rev. Res. **2**, 043416 (2020).

[45] H. M. Abdullah, D. R. da Costa, H. Bahlouli, A. Chaves, F. M. Peeters, and B. Van Duppen, *Electron Collimation at van Der Waals Domain Walls in Bilayer Graphene*, Phys. Rev. B **100**, 045137 (2019).

[46] W.-Y. He, Z.-D. Chu, and L. He, *Chiral Tunneling in a Twisted Graphene Bilayer*, Phys. Rev. Lett. **111**, 066803 (2013).

[47] A. K. Geim, *Graphene: Status and Prospects*, Science **324**, 1530 (2009).

[48] Y. Cao, V. Fatemi, S. Fang, K. Watanabe, T. Taniguchi, E. Kaxiras, and P. Jarillo-Herrero, *Unconventional Superconductivity in Magic-Angle Graphene Superlattices*, Nature **556**, 7699 (2018).

[49]     C. R. Dean, A. F. Young, I. Meric, C. Lee, L. Wang, S. Sorgenfrei, K. Watanabe, T. Taniguchi, P. Kim, and K. L. Shepard, *Boron Nitride Substrates for High-Quality Graphene Electronics*, Nat. Nanotechnol. **5**, 722 (2010).

[50]     L. Wang, I. Meric, P. Y. Huang, Q. Gao, Y. Gao, H. Tran, T. Taniguchi, K. Watanabe, L. M. Campos, and D. A. Muller, *One-Dimensional Electrical Contact to a Two-Dimensional Material*, Science **342**, 614 (2013).

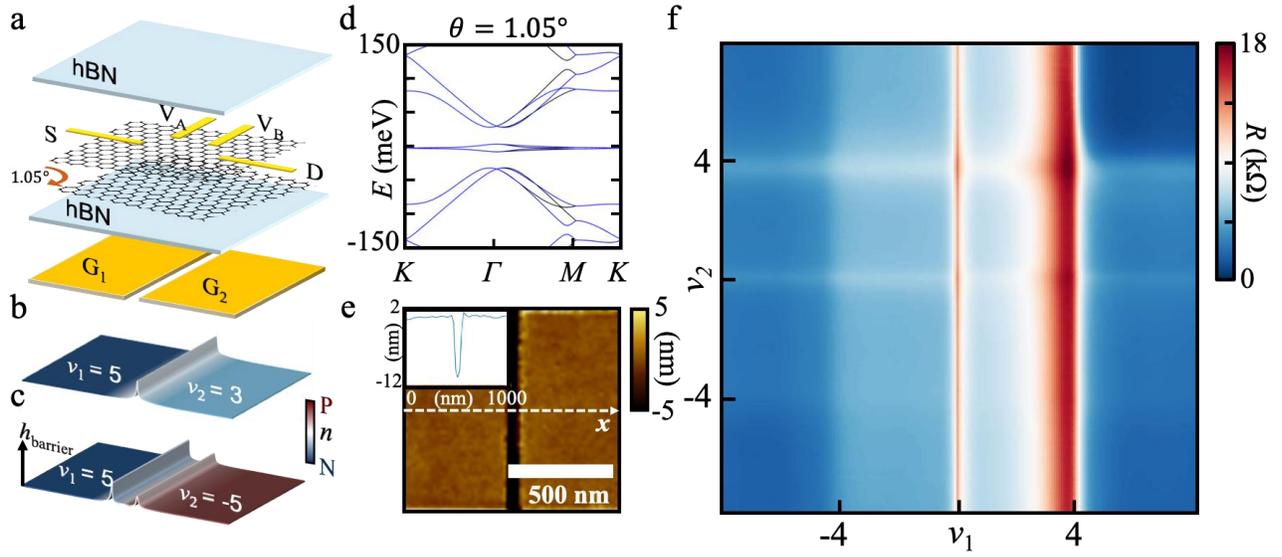

**Figure 1. Gate-defined Moiré Barriers in tBLG.** (a) Schematic image of a moiré junction device. Two pieces of MLG are consecutively picked up with a relative twist angle of 1.05°, then transferred on top of prepatterned local bottom gates (G$_1$ and G$_2$) with 100 nm gate separation. 1D edge contact are subsequently deposited to serve as source (S), drain (D), and voltage probes (V$_A$, V$_B$) for transport measurement. (b) Bottom: Schematic of the carrier density distribution in the device when carrier density is tuned to be $\nu_1 = 5$ ($\nu_2 = 3$) for the region on top of G$_1$ (G$_2$). Carrier density changes from $\nu = \nu_1$ to $\nu = \nu_2$ in regions above the gate separation, a narrow section of which reaches $\nu = 4$ band insulator states and serve as a narrow tunnel barrier, which we refer to as $\nu = 4$ moiré barrier. (c) Schematic of the carrier density distribution in the device when carrier density is tuned to be $\nu_1 = 5$ and $\nu_2 = -5$. Three narrow barriers (corresponding to $\nu = -4$, $\nu = 0$ and $\nu = 4$) are present in series at the junction. (d) Band structure of tBLG with twist angle of 1.05°. Blue and black come from K and K' valley respectively. (e) Atomic force microscope (AFM) topography of local bottom gates. Inset: 1D cuts along the white dashed line demonstrating atomic flatness, essential for a homogeneous moiré barrier. (f) Measured 4-probe resistance as a function of $\nu_1$ and $\nu_2$. Local high resistance is observed along three horizontal and vertical lines, when either one of the two regions reaches $\nu = -4, 0$ or $4$.

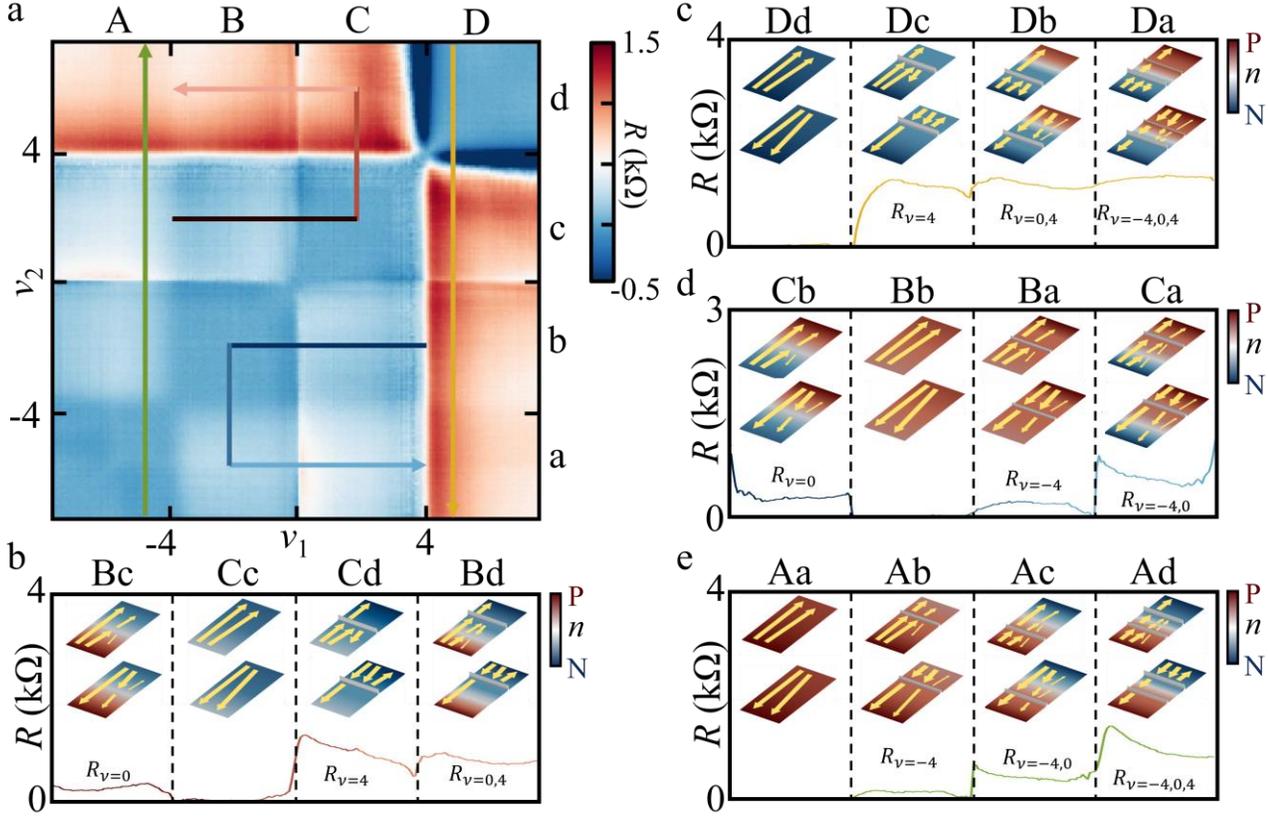

**Figure 2. Electron Collimation via Moiré Barriers.** (a) Extracted junction resistance $R_J$ as a function of filling factors $\nu_1$ and $\nu_2$. 16 domains separated by $\nu_1, \nu_2 = -4, 0, 4$ lines are clearly visible. Column number A, B, C and D and row number a, b, c and d are used to label different $\nu_1$ and $\nu_2$ sections, respectively. (b) 1D cut along the red path in (a), where the junction contains tunnel barriers of Bc:$[\nu = 0] \to$ Cc:[none] $\to$ Cd:$[\nu = 4] \to$ Bd:$[\nu = 0, 4]$. The observation of $R_{\nu=0} + R_{\nu=4} > R_{\nu=0,4} \approx R_{\nu=4}$ implies nearly perfect (some) collimation from the $\nu = 4$ moiré barrier ($\nu = 0$ pn-barrier). Inset: schematic image of the charge carriers passing through and being reflected by the junction. The thickness of the arrows indicates the number of charge carriers. The color bar on the right side corresponds to the carrier density. (c) 1D cut along the yellow path in (a) with junction consisting of barriers from Dd:[none] $\to$ Dc:$[\nu = 4] \to$ Db: $\to [\nu = 0, 4] \to$ Da:$[\nu = -4, 0, 4]$. (d) 1D cut along the blue path in (a), consisting of barriers from Cb:$[\nu = 0] \to$ Bb:[none] $\to$ Ba: $[\nu = -4] \to$ Ca:$[\nu = -4, 0]$. (e) 1D cut along the green path in (a), consisting of barriers from: Aa:[none] $\to$ Ab:$[\nu = -4] \to$ Ac: $\to [\nu = -4, 0] \to$ Ad:$[\nu = -4, 0, 4]$.

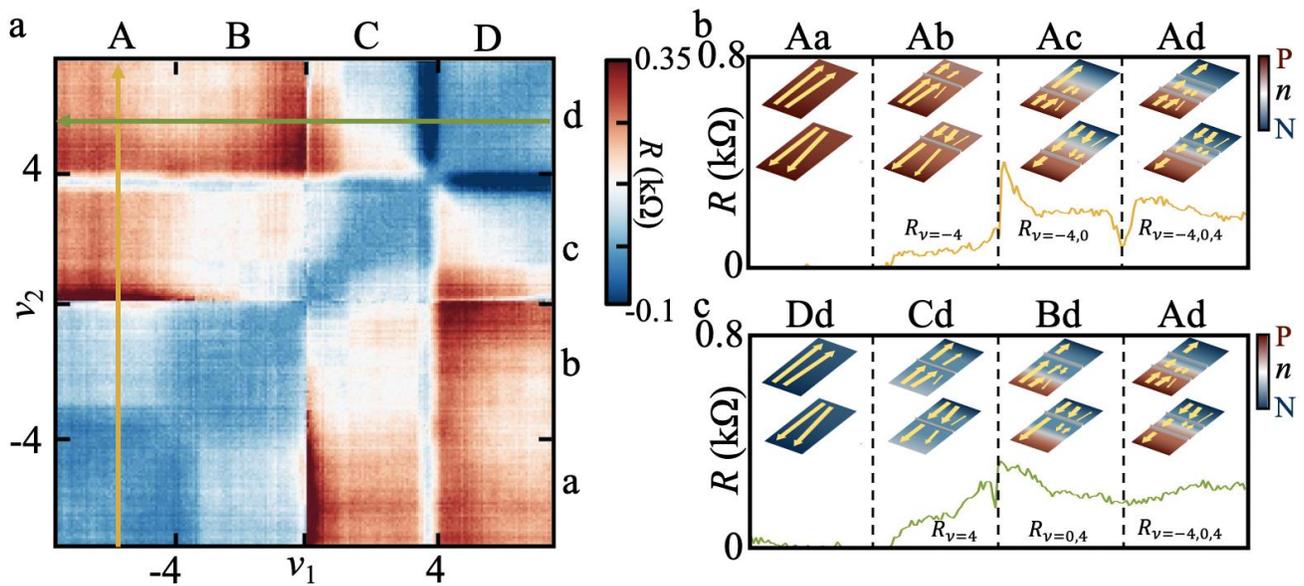

**Figure 3. Characterization of Electron Collimation from the Second Moiré Junction.** (a) Junction resistance as a function of $\nu_1$ and $\nu_2$. The $\nu_1, \nu_2 = -4, 0, 4$ local resistance peaks separating the 16 sections are found at nearly the exact same carrier densities as that of junction 1, showing that the overall twist-angle has not migrated over the 1 μm separation between the two devices. (b)(c) 1D cuts through Column A (yellow) and row d (green) imply that neither of the two ($\nu = -4$ and $\nu = 4$) moiré barriers is enough to reach a high level of collimation by itself. However, the junction resistances with two consecutive barriers (Ac: $\nu = 0, 4$ or Bd: $\nu = -4, 0$), are more similar to that of three consecutive barriers (Ad: $\nu = -4, 0, 4$). This shows that effective collimation can be achieved with two tunnel barriers in the junction, and the addition of the third (final) barrier no longer helps.

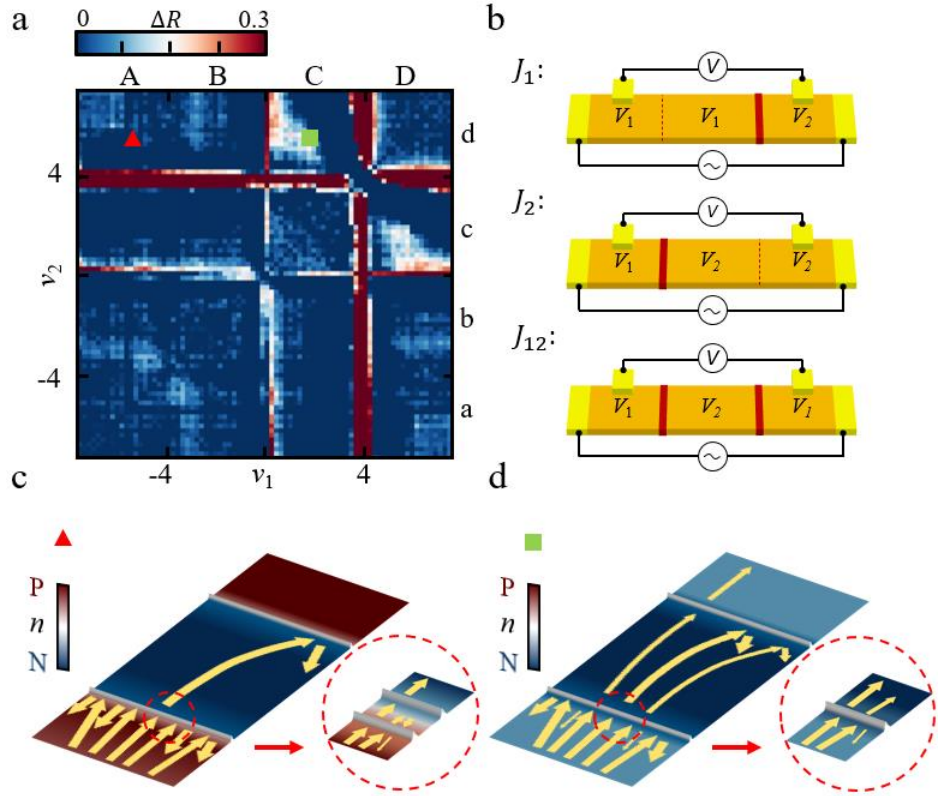

**Figure 4. Characterization of Long-distance Collimation Efficiency in tBLG.** (a) $\Delta R = R_{J_1} + R_{J_2} - R_{J_{12}}$ as a function of $\nu_1$ and $\nu_2$, with $R_{J_1}, R_{J_2}, R_{J_{12}}$ measured with the (b) same circuits but different gate configurations. $\Delta R = 0$ corresponding to the trivial scenario of two resistors in series. A positive $\Delta R$ is observed in section Cd (green square) and Dc, when both junctions consist of only $\nu = 4$ moiré barriers, demonstrating a signature of electron collimation over 1 µm distance in tBLG. (c) Schematic of electron flow at the limit of nearly perfect collimation by each junction, where any electron other than the one incident strictly perpendicular to the barrier is reflected (inset). The curvature of the collimated electron path over the 1 µm distance results in reflections of electrons with finite incident angle, instead of 100% transmission, eliminating the expected logic between the two collimators. (d) Schematic of electron flow when both junctions consist of only $\nu = 4$ moiré barriers. Insets: Zoom-in on the electron trajectory through $J_2$. Electron through $J_2$ has a narrow but still finite angle distribution, permitting electron paths through $J_2$ with small incident angle. A selection of those electrons will incident perpendicularly to $J_1$ after traveling a curved ballistic path, successfully transmitting through $J_1$ with ease and contributing to $\Delta R$.

# Electron Collimation in Twisted Bilayer Graphene via Gate-defined Moiré Barriers

Wei Ren[1], Xi Zhang[1], Ziyan Zhu[2], Moosa Khan[1], Kenji Watanabe[3], Takashi Taniguchi[4], Efthimios Kaxiras[5,6], Mitchell Luskin[7], Ke Wang[1*]

[1]*School of Physics and Astronomy, University of Minnesota, Minneapolis, Minnesota 55455, USA*

[2]*Stanford Institute for Materials and Energy Sciences, SLAC National Accelerator Laboratory, Menlo Park, CA 94025, USA*

[3]*Research Center for Electronic and Optical Materials, National Institute for Materials Science, 1-1 Namiki, Tsukuba 305-0044, Japan*

[4]*Research Center for Materials Nanoarchitectonics, National Institute for Materials Science, 1-1 Namiki, Tsukuba 305-0044, Japan*

[5]*Department of Physics, Harvard University, Cambridge, Massachusetts 02138, USA*

[6]*John A. Paulson School of Engineering and Applied Sciences, Harvard University, Cambridge, Massachusetts 02138, USA*

[7]*School of Mathematics, University of Minnesota, Minneapolis, Minnesota 55455, USA*

## S1. Characterization of Moiré Barrier Width via Measured Transport Resistance

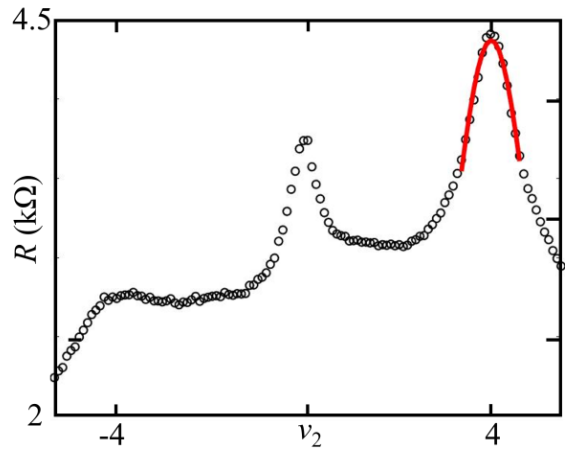

**Figure S1. Fitting of the resistance peak at a moiré barrier.** Black circles represent the original data while the red curve is the peak fitting result. The half-height-width of the peak is estimated as $\Delta\nu = 0.96$. The width of the moiré barrier is characterized as ~ 9.6 nm.

The width of the moiré barrier can be characterized through the measured transport resistance. Figure S1 shows a 1D resistance line cut along $\nu_1 = -5$ in the range of $\Delta\nu_2 = 10$, where an obvious resistance peak can be observed at $\nu_2 = 4$, which corresponds to the moiré barrier at the n-side. The red curve represents the peak fitting result, agreeing with the original data points which are indicated by the black circles. The width at half-height of the peak is estimated as $\Delta\nu = 0.96$. As the size of the gap between the gates $d$ is ~100 nm, the width of the moiré barrier can be characterized by the equation:

$$w = d\frac{\Delta\nu}{\Delta\nu_2} = 100 \text{ nm} \times \frac{0.96}{10} = 9.6 \text{ nm}.$$

## S2. Extraction of Junction Resistance via the Symmetrization Method

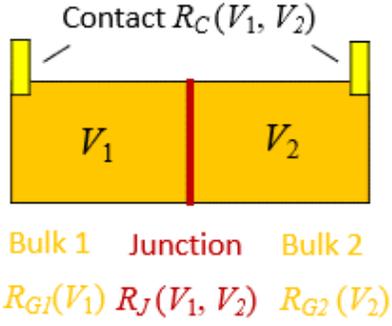

**Figure S2. Schematic Image of the Device.** The two bright yellow bars represent metal contacts. The dark yellow rectangles represent bulk regions of the tBLG devices, independently controlled by the two local bottom gate voltages $V_1$ and $V_2$. The red region represents the moiré and/or pn barrier.

In a ballistic device, the resistance of the junction characterizes the backscattering at the junction, while the bulk resistance of the graphene is limited by ballistic quantum conductance [1,2]. To characterize the collimation effect of a junction (either a conventional pn-barrier, a moiré barrier, or a combination of them), we adopt the following symmetrization method to isolate the junction resistance from the previously shown 4-probe resistance. The schematic image of a device with two gate-tunable regions is shown in Figure S2a. The contact resistance, the resistance of bulk graphene directly on top of the local bottom gates and the junction resistance are denoted by $R_C$, $R_{G1}$, $R_{G2}$, and $R_J$ respectively. $R_{G1}(R_{G2})$ is a function of $V_1(V_2)$ only, while $R_J$ is function of both $V_1$ and $V_2$. Note that $R_C$ comes from the process where electrons overcome the Schottky barrier between graphene and the metal voltage probes. Although the Schottky barrier [3] is determined by the properties of the materials, i.e., graphene and Cr/Pd/Au metal contacts, the change in the carrier density in graphene may still weakly change the contact resistance. So, here, we denote $R_J$ as a function of both $V_1$ and $V_2$. Besides, we expect the contact and junction resistance are symmetric functions of $V_1$ and $V_2$, i.e.,

$$R_C(V_1, V_2) = R_C(V_2, V_1),$$

$$R_J(V_1, V_2) = R_J(V_2, V_1).$$

When the left and right gate voltage are tuned to be $V_1$ and $V_2$, the total resistance of the device can be expanded into the summation of four terms:

$$R_T(V_1, V_2) = R_C(V_1, V_2) + R_{G1}(V_1) + R_{G2}(V_2) + R_J(V_1, V_2).$$

Similarly, the resistance of the device can be expanded as the following when the two gate voltages are at $(V_2, V_1)$, $(V_1, V_1)$ and $(V_2, V_2)$:

$$R_T(V_2, V_1) = R_C(V_2, V_1) + R_{G1}(V_2) + R_{G2}(V_1) + R_J(V_2, V_1),$$

$$R_T(V_1, V_1) = R_C(V_1, V_1) + R_{G1}(V_1) + R_{G2}(V_1) + R_J(V_1, V_1),$$

$$R_T(V_2, V_2) = R_C(V_2, V_2) + R_{G1}(V_2) + R_{G2}(V_2) + R_J(V_2, V_2),$$

where $R_J(V_1, V_1) = R_J(V_2, V_2) = 0$ since there is no junction formed when the device is uniformly gated. Therefore, the junction resistance is extracted to be:

$$R_J(V_1, V_2) = \frac{1}{2}(R_T(V_1, V_2) + R_T(V_2, V_1) - R_T(V_1, V_1) - R_T(V_2, V_2)).$$

## S3. Charactering the Collimation Efficiency of a Junction

With the junction resistance being extracted, the collimation efficiency is further characterized by analyzing the logic between two consecutive junctions ($J_1$ and $J_2$), i.e., the relationship between the symmetrized resistance $R_{J1}$, $R_{J2}$, and $R_{J12}$. Here, $R_{J1}$ ($R_{J2}$) is the junction resistance when only $J_1$ ($J_2$) is defined, while $R_{J12}$ is the total junction resistance when both $J_1$ and $J_2$ are simultaneously defined and connected in series. $J_1$ and $J_2$ can be considered as either (1) two barriers (either conventional pn-barriers or moiré barriers) coexsisting within the same gap (~100 nm wide) between two local bottom gates, which is the case discussed in Figure 2 and Fig. 3, or (2) two junctions that are 1 μm away as discussed in Figure 4 in the main manuscript.

In the limit of perfect collimation by junction $J_1$, charge carriers will experience no reflection at $J_2$ whether $J_2$ is established or not, leading to $R_{J12} = R_{J1}$; While in the limit of complete absence of collimation, we expect $R_{J12} = R_{J1} + R_{J2}$, similar to two resistors in series. In previous state-of-the-art pn-junctions, it is found that $R_{J1} + R_{J2} > R_{J12} > \max(R_{J1}, R_{J2})$ with at least 25% difference between neighboring terms [4]. This corresponds to about 30% collimation efficiency, meaning that more than half of the charge carriers stay uncollimated.

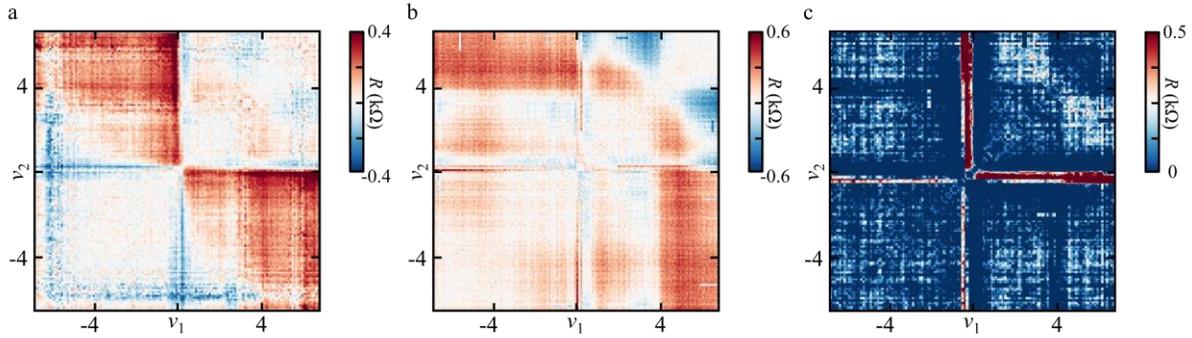

**Figure S3. Additional Devices Data.** (a) Junction resistance as a function of $v_1$ and $v_2$ at one junction in the additional device. (b) Junction resistance as a function of $v_1$ and $v_2$ at another junction in the same additional device. (c) $\Delta R = R_{J_1} + R_{J_2} - R_{J_{12}}$ as a function of $v_1$ and $v_2$.

## S4. Transport Signature from the Additional Device

In addition to the tBLG device that was presented in Figure 2 and 3 in the main manuscript, a similar transport signal can be observed in the control device, due to the existence of a moiré barrier. As Figure S3(a)-(b) show, a moiré barrier is formed on the n-side in an additional device. However, the transport feature as well as the strength of the moiré barrier (in terms of resistance) are not strong enough as the data that is presented in Figure 2 or Figure 3, which could be attributed to the large angle inhomogenity. This is consistent with the broader transition region on the n-side in Figure S3(a) and (b), indicating varied moiré lengths in the device. Similar to Figure 2, the electron-hole symmetry-breaking is also found in the additional devices, where the moiré barrier is not even formed on the p-side in this additional device. This might be due to the absence of the moiré gap on the p-side corresponding to their specific twist angles. Similar to Figure 4(a) in the main paper, positive $\Delta R$ is observed in the NN regime in Figure S3(c), demonstrating the signature of electron collimation in the tBLG device.

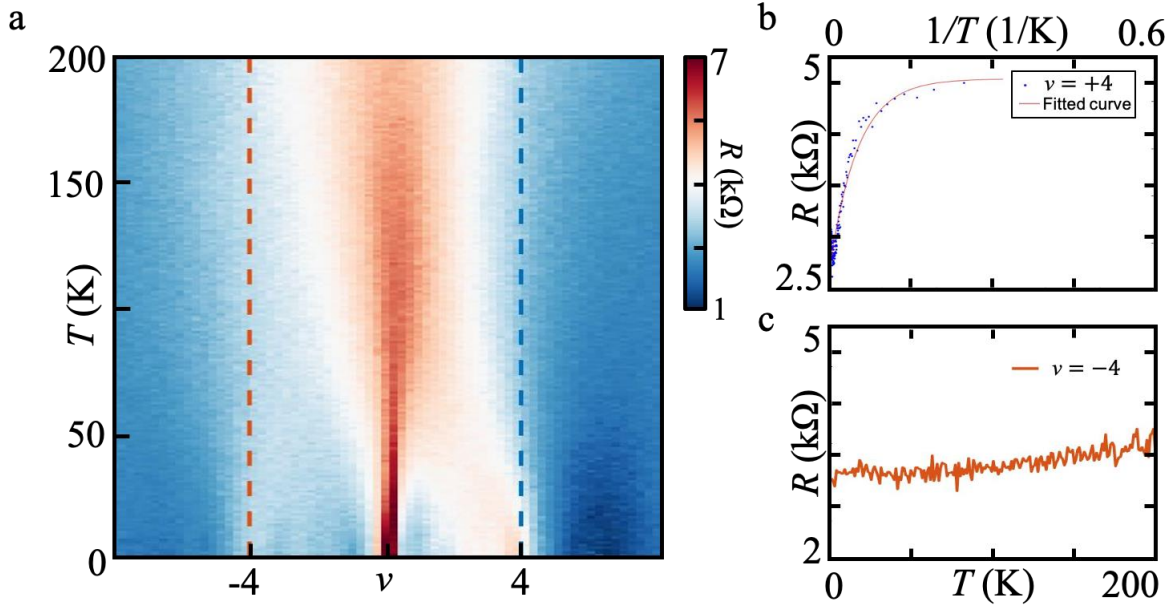

**Figure S4. Temperature Dependence of R.** (a) $R$ of Device 1 as a function of moiré filling factors $v$ and temperature $T$. (b) 1D cuts of (a) along $v = +4$ but plotted as a function of $1/T$. An exponential fit is overlaid on top of the data points. (c) 1D cuts of (a) along $v = -4$, where almost no temperature dependence is observed.

## S5. Characterization on Moiré Band Gaps via Temperature Dependence

The temperature dependence of 4-probe longitudinal resistance can be used to determine the size of the moiré band gaps at $v = \pm 4$. Note that both local gates are tuned at the same voltage in this measurement to measure the global transport signature of the tBLG devices (no presence of any moiré barrier). The following equation is used to fit the resistance measured at $v = \pm 4$ as a function of temperature $T$ [5,6]:

$$R \propto e^{\frac{-E_{gap}}{2k_B T}}, \tag{1}$$

where $E_{gap}$ is the size of the band gaps, and $k_B$ is the Boltzmann factor. A temperature dependent scan at various filling factors is shown in Figure S4a. A vertical cut is taken at $v = +4$ and an exponential fit following Equation (1) is applied to this cut (Figure S4b). The fitting yields a gap energy of $3.36 \pm 0.44$ meV, which is smaller than the calculated band insulator gap (~23 meV). This indicates the inhomogeneity in the sample region across the two bottom gates (at least 1 µm × 2 µm) has a considerable effect. Possible causes may include variations in local twist angle, non-uniform doping of tBLG across the sample region, the formation of conductive domain walls resulting from lattice reconstruction (particularly notable when the twist angle approaches the critical angle, ~ 1°, for significant lattice reconstruction), and other factors. Despite the potential presence of inhomogeneity over long distances, the band insulator gap at the junction may remain well-defined, given the relatively small length scale of the junction region (~100 nm). The collimation effect resulting from a clearly defined moiré barrier remains observable in the measurements. For $v = -4$, the resistance remains nearly constant as a function of temperature (Figure S4c), suggesting a vanishing band gap. The uncertainty on the exponential fit is larger than the extracted energy gap, so

Equation (1) is not applicable anymore. The result matches with what we observe in the main manuscript (Figure 2a) that a larger moiré band gap (in this case it is the band gap present at $\nu = +4$) can create a more effective Klein tunneling barrier and thus facilitates the collimation process.

**S6. Band Structure Calculation**

We perform a band structure calculation of the twisted bilayer graphene using a low-energy continuum model, incorporating the out-of-plane relaxation [7,8]. In the low-energy limit, the intralayer term can be approximated by rotated linear Dirac Hamiltonians:

$$H_D^1(\mathbf{q}^{(1)}) = -v_F \begin{bmatrix} 0 & e^{i\theta/2}\mathbf{q}_+^{(1)} \\ e^{-i\theta/2}\mathbf{q}_-^{(1)} & 0 \end{bmatrix},$$

$$H_D^3(\mathbf{q}^{(2)}) = -v_F \begin{bmatrix} 0 & e^{-i\theta/2}\mathbf{q}_+^{(2)} \\ e^{i\theta/2}\mathbf{q}_-^{(2)} & 0 \end{bmatrix}, \quad (2)$$

where $\mathbf{q}^{(\ell)}$ is the momentum degree of freedom in layer $\ell$, $\mathbf{q}_\pm^{(l)} = \mathbf{q}_x^{(l)} \pm \mathbf{q}_y^{(l)}$, and $v_F$ is the Fermi velocity of monolayer graphene which we take to be $v_F = 0.8 \times 10^6$ m/s from the DFT calculated value. For the interlayer coupling, we keep the nearest neighbor coupling in momentum space:

$$T(\mathbf{q}^{(1)}, \mathbf{q}^{(2)}) = \sum_{n=1}^{3} T_{\alpha\beta}^{\mathbf{q}_n} \delta_{\mathbf{q}^{(1)} - \mathbf{q}^{(2)}, -\mathbf{q}_n}, \quad (3)$$

where $\alpha\beta$ are sublattice degrees of freedom, $\mathbf{q}_1 = K_{L_1} - K_{L_2}$, $\mathbf{q}_2 = \mathcal{R}^{-1}\left(\frac{2\pi}{3}\right)\mathbf{q}_1$, $\mathbf{q}_3 = \mathcal{R}\left(\frac{2\pi}{3}\right)\mathbf{q}_1$ and $\mathcal{R}(\theta)$ is the counterclockwise rotation matrix by $\theta$, $K_{L_\ell}$ is the Dirac point of layer $\ell$. We take into account the out-of-plane relaxation by letting $t_{AA}^{ij} = t_{BB}^{ij} = \omega_0 = 0.07$ eV and $t_{AB}^{ij} = t_{BA}^{ij} = \omega_1 = 0.1$ eV due to the strengthened interaction between AB/BA sites from relaxation

$$T^{\mathbf{q}_1} = \begin{bmatrix} \omega_0 & \omega_1 \\ \omega_1 & \omega_0 \end{bmatrix}, T^{\mathbf{q}_2} = \begin{bmatrix} \omega_0 & \omega_1\bar{\phi} \\ \omega_1\phi & \omega_0 \end{bmatrix}, T^{\mathbf{q}_3} = \begin{bmatrix} \omega_0 & \omega_1\phi \\ \omega_1\bar{\phi} & \omega_0 \end{bmatrix}, \quad (4)$$

where $\phi = e^{\frac{i2\pi}{3}}$ and $\bar{\phi} = e^{-\frac{i2\pi}{3}}$.

**References**

[1] R. A. Jalabert, H. U. Baranger, and A. D. Stone, *Conductance Fluctuations in the Ballistic Regime: A Probe of Quantum Chaos?*, Phys. Rev. Lett. **65**, 2442 (1990).
[2] G. Kirczenow, *Theory of the Conductance of Ballistic Quantum Channels*, Solid State Communications **68**, 715 (1988).
[3] R. T. Tung, *The Physics and Chemistry of the Schottky Barrier Height*, Applied Physics Reviews **1**, 011304 (2014).

[4] K. Wang, M. M. Elahi, L. Wang, K. M. Habib, T. Taniguchi, K. Watanabe, J. Hone, A. W. Ghosh, G.-H. Lee, and P. Kim, *Graphene Transistor Based on Tunable Dirac Fermion Optics*, Proceedings of the National Academy of Sciences **116**, 6575 (2019).
[5] G. Chen et al., *Evidence of a Gate-Tunable Mott Insulator in a Trilayer Graphene Moiré Superlattice*, Nat. Phys. **15**, 3 (2019).
[6] C. Ma et al., *Moiré Band Topology in Twisted Bilayer Graphene*, Nano Lett. **20**, 6076 (2020).
[7] R. Bistritzer and A. H. MacDonald, *Moiré Bands in Twisted Double-Layer Graphene*, Proceedings of the National Academy of Sciences **108**, 12233 (2011).
[8] N. N. T. Nam and M. Koshino, *Lattice Relaxation and Energy Band Modulation in Twisted Bilayer Graphene*, Phys. Rev. B **96**, 075311 (2017).